\documentclass[prl,twocolumn,superscriptaddress,showpacs,preprintnumbers]{revtex4}
\usepackage{graphicx,epsfig}
\begin{document}

\def\beqra{\begin{eqnarray}} \def\eeqra{\end{eqnarray}}
\def\beqast{\begin{eqnarray*}} \def\eeqast{\end{eqnarray*}}
\def\beq{\begin{equation}}      \def\eeq{\end{equation}}
\def\be{\begin{enumerate}}   \def\ee{\end{enumerate}}

\def\gam{\gamma}
\def\Gam{\Gamma}
\def\la{\lambda}
\def\eps{\epsilon}
\def\La{\Lambda}
\def\si{\sigma}
\def\Si{\Sigma}
\def\al{\alpha}
\def\Tha{\Theta}
\def\tha{\theta}
\def\vphi{\varphi}
\def\del{\delta}
\def\Del{\Delta}
\def\ab{\alpha\beta}
\def\om{\omega}
\def\Om{\Omega}
\def\mn{\mu\nu}
\def\mun{^{\mu}{}_{\nu}}
\def\kap{\kappa}
\def\rsi{\rho\sigma}
\def\beal{\beta\alpha}
\def\til{\tilde}
\def\rta{\rightarrow}
\def\eqv{\equiv}
\def\nab{\nabla}
\def\pa{\partial}
\def\sit{\tilde\sigma}
\def\ul{\underline}
\def\indt{\parindent2.5em}
\def\nd{\noindent}
\def\rsi{\rho\sigma}
\def\beal{\beta\alpha}
\def\caa{{\cal A}}
\def\cb{{\cal B}}
\def\cac{{\cal C}}
\def\cd{{\cal D}}
\def\ce{{\cal E}}
\def\cf{{\cal F}}
\def\cg{{\cal G}}
\def\cah{{\cal H}}
\def\ci{{\cal I}}
\def\cj{{\cal{J}}}
\def\ck{{\cal K}}
\def\cl{{\cal L}}
\def\cm{{\cal M}}
\def\cn{{\cal N}}
\def\cO{{\cal O}}
\def\cp{{\cal P}}
\def\car{{\cal R}}
\def\cs{{\cal S}}
\def\ct{{\cal{T}}}
\def\cu{{\cal{U}}}
\def\cv{{\cal{V}}}
\def\cw{{\cal{W}}}
\def\cx{{\cal{X}}}
\def\cy{{\cal{Y}}}
\def\cz{{\cal{Z}}}
\def\asymptotic{{_{\stackrel{\displaystyle\longrightarrow}
{x\rightarrow\pm\infty}}\,\, }} 
\def\asymptext{\raisebox{.6ex}{${_{\stackrel{\displaystyle\longrightarrow}
{x\rightarrow\pm\infty}}\,\, }$}} 
\def\asymptoticp{{_{\stackrel{\displaystyle\longrightarrow}
{x\rightarrow +\infty}}\,\, }} 
\def\asymptoticm{{_{\stackrel{\displaystyle\longrightarrow}
{x\rightarrow -\infty}}\,\, }} 

\def\raisenot{\raise .5mm\hbox{/}}
\def\nota{\ \hbox{{$a$}\kern-.49em\hbox{/}}}
\def\notA{\hbox{{$A$}\kern-.54em\hbox{\raisenot}}}
\def\notb{\ \hbox{{$b$}\kern-.47em\hbox{/}}}
\def\notB{\ \hbox{{$B$}\kern-.60em\hbox{\raisenot}}}
\def\notc{\ \hbox{{$c$}\kern-.45em\hbox{/}}}
\def\notd{\ \hbox{{$d$}\kern-.53em\hbox{/}}}
\def\notbd{\ \hbox{{$D$}\kern-.61em\hbox{\raisenot}}} 
\def\note{\ \hbox{{$e$}\kern-.47em\hbox{/}}}
\def\notk{\ \hbox{{$k$}\kern-.51em\hbox{/}}}
\def\notp{\ \hbox{{$p$}\kern-.43em\hbox{/}}}
\def\notq{\ \hbox{{$q$}\kern-.47em\hbox{/}}}
\def\notW{\ \hbox{{$W$}\kern-.75em\hbox{\raisenot}}}
\def\notz{\ \hbox{{$Z$}\kern-.61em\hbox{\raisenot}}}
\def\notpa{\hbox{{$\partial$}\kern-.54em\hbox{\raisenot}}}

\def\fo{\hbox{{1}\kern-.25em\hbox{l}}}  
\def\rf#1{$^{#1}$}
\def\bx{\Box}
\def\tr{{\rm Tr}}
\def\rmtr{{\rm tr}}
\def\dgg{\dagger}
\def\lag{\langle}
\def\rag{\rangle}
\def\bmid{\big|}
\def\vlap{\overrightarrow{\La p}} 
\def\lrta{\longrightarrow} \def\lrar{\raisebox{.8ex}{$\longrightarrow$}}
\def\ON{{\cal O}(N)}
\def\UN{{\cal U}(N)}
\def\bdPh{\mbox{\boldmath{$\dot{\!\Phi}$}}}
\def\bPh{\mbox{\boldmath{$\Phi$}}}
\def\bPhs{\bPh^2}
\def\sef{S_{eff}[\sigma,\pi]}
\def\sigx{\sigma(x)}
\def\pix{\pi(x)}
\def\bph{\mbox{\boldmath{$\phi$}}}
\def\bphs{\bph^2}
\def\ex{\BM{x}}
\def\exs{\ex^2}
\def\xdot{\dot{\!\ex}}
\def\y{\BM{y}}
\def\ys{\y^2}
\def\ydot{\dot{\!\y}}
\def\pat{\pa_t}
\def\pax{\pa_x}


\title{A Universal Scaling Theory for Complexity of Analog Computation}

\author{Yaniv S. Avizrats}
\affiliation{Department of Physics, Technion, Haifa 32000, Israel.}
\author{Shmuel Fishman}
\affiliation{Department of Physics, Technion, Haifa 32000, Israel.}
\author{Joshua Feinberg}
\affiliation{Department of Physics, University of Haifa at Oranim,
Tivon 36006, Israel.}
\affiliation{Department of Physics, Technion, Haifa 32000, Israel.}


\begin{abstract}
We discuss the computational complexity of solving linear programming 
problems by means of an analog computer. The latter is modeled by a dynamical
system which converges to the optimal vertex solution. We analyze various 
probability ensembles of linear programming problems. For each one of these 
we obtain numerically the probability distribution functions of 
certain quantities which measure the complexity. Remarkably, in the 
asymptotic limit of very large problems, each of these probability 
distribution functions reduces to a universal scaling function, depending 
on a single scaling variable and independent of the details of its parent 
probability ensemble. These functions are reminiscent of the scaling 
functions familiar in the theory of phase transitions. The results reported 
here extend analytical and numerical results obtained recently for the 
Gaussian ensemble.
\end{abstract}


\maketitle


During the past two decades or so, physicists have been applying methods
of statistical physics in studying hard combinatorial optimization problems
of computer science \cite{mpv}. Physical methods, borrowed from statistical 
physics and the theory of dynamical systems (DS) \cite{ott}, have been applied 
recently in studying an easier computer science problem, namely, the computational 
complexity of an analog computation algorithm which solves linear programming 
(LP) problems \cite{bffs, pla}. In this Letter we demonstrate the robustness
and universality of the results of \cite{bffs, pla}.

To put things in broader context, we remark that analog computers are 
ubiquitious computational tools, alongside with their predominating digital 
counterparts. The most relevant examples of analog computers are VLSI devices 
implementing neural networks \cite{Hertznn-optimwang}, or neuromorphic systems 
\cite{mead}, whose structure is directly motivated by the workings of the brain.
Various processes taking place in living cells can be considered as analog 
computation \cite{bio} as well.

Linear programming is a P-complete problem \cite{Papadimitriou}, i.e.
it is representative of all problems that can be solved in polynomial time.
The {\it standard form} of LP is to find 
\begin{equation}
\label{standard}
\max \{ c^T x ~:~ x\in I\!\! R ^{n}, A x =  b,x \geq 0  \}
\end{equation}
where $c \in I\!\! R^n, b \in I\!\! R^m, A \in I\!\! R^{m \times n}$ and $m\leq n$.
The set generated by the constraints in (\ref{standard}) is a polyheder.
If a bounded optimal solution exists, it is obtained at one of its vertices.
The vector defining this optimal vertex can be decomposed (in an appropriate 
basis) in the form $x=(x_{{\cal N}},x_{{\cal B}})$ where $x_{\cal N} = 0$ is 
an $n-m$ component vector, while $x_{\cal B} =  B^{-1}  b \geq  0$ is an $m$ 
component vector, and $B$ is the $m \times m$ matrix whose columns are
the columns of $A$ with indices identical to the ones of $x_{\cal B}$.
Similarly, we decompose $A=(N,B)$, where $N$ is an $(n-m)\times m$ matrix.

A DS whose flow converges to the optimal vertex, introduced by Faybusovich 
\cite{faybusovich}, will be studied here. Its flow $\frac{dx}{dt}=F(x)$
is given in terms of the vector field 
\begin{equation} \label{field}
F(x)=[X - X A^{T}(A X A^{T})^{-1} A X]\: c\; ,
\end{equation}
where $X$ is the diagonal matrix $\mbox{Diag}(x_1 \dots x_n)$.
Geometrically, $F$ is the projection of the gradient of 
the cost function $c^T x$ onto the constraint set, relative to a
Riemannian metric, which enforces the positivity constraints $x\geq 0$
\cite{faybusovich}. 

This DS, as it evolves in its continuous phase space in continuous time,  
models the analog computer in question, which solves the given LP problem. 
Other dynamical systems are known (also described by ordinary differential 
equations) that are used to solve various computational problems 
\cite{brockett,faybusovich,helmke-moore}. Thus, a large set of analytical tools 
and physical intuition, developed for dynamical systems, turns out to be 
applicable to the analysis of analog computers.

In contrast, the evolution of a digital computer is described by a
dynamical system, discrete both in its phase space and in time. 
Consequently, the standard theory of computation and computational 
complexity \cite{Papadimitriou} deals with computation in discrete time
and phase space, and is inadequate for the description of
analog computers. The analysis of computation by analog
devices requiers a theory that is valid in continuous time and phase
space.  

Since the systems in question are physical systems, the computation time is 
the time required for a system to reach the vicinity of an attractor 
(a stable fixed point in the present work) combined with the time
required to verify that it indeed reached this vicinity.
This time is the elapsed time measured by a clock, contrary
to standard computation theory, where it is the number of discrete 
steps.

In our model we assume we have a physical implementation of the flow 
equation. Thus, the vector field $F$ need not be computed, 
and the computation time is determined by the convergence time to the 
attracting fixed point. In other words, the time of flow to the vicinity
of the attractor is a good measure of complexity, namely the computational 
effort, for the class of continuous dynamical systems introduced above 
\cite{dds}. 

In this work, following \cite{bffs, pla} (and in a manner similar to
\cite{lpsmaleTodd-models,shamir}), the complexity will be evaluated 
probabilistically for an ensemble of LP problems. In this way the worst case 
scenarios, studied traditionally in computer science, will be 
ignored, since their probability measure vanishes. 

The main result of \cite{bffs, pla}, in which LP problems were drawn from 
the Gaussian distribution of the parameters of $F$ (namely, the 
constraints and cost function in (\ref{standard})), was that the 
distribution functions of various
quantities that characterize the computational complexity, were found to be 
scaling functions in the limit of LP problems of large size. In particular,
it was found that those distribution functions depended on the various 
parameters only via specific combinations, namely, the scaling variables.
Such behavior is analogous to the situation found for the
central limit theorem, for critical phenomena \cite{wilson}
and for Anderson localization \cite{anderson}, in spite of the very different 
nature of these problems. It was demonstrated in \cite{bffs, pla} how for 
the implementation of the LP problem on a physical device, methods used in
theoretical physics enable to describe the distribution
of computation times in a simple and physically transparent form.
Based on experience with certain universality properties of
rectangular and chiral random matrix models,
it was conjectured in \cite{bffs, pla} that some universality for 
computational problems should be expected and should be explored. That is, 
the scaling properties that were found for the Gaussian distributions should 
hold also for other distributions. In particular, some specific 
questions were raised in \cite{bffs, pla}:
Is the Gaussian nature of the ensemble unimportant? 
Are there universality classes \cite{wilson}
of analog computational problems, and if they exist, what are they? 
In the present work we extend the earlier analysis \cite{bffs, pla} of the 
Gaussian distribution to other probability distributions, and demonstrate 
{\em numerically} that the distribution functions of various quantities that 
characterize the computational complexity of the analog computer, which 
solves LP problems, are indeed {\em universal} scaling functions. They 
depend upon the original probability ensemble of inputs only via the scaling 
variables, that are proportional to the ones found for the Gaussian 
distribution.

The distribution of constraints and cost function of the LP problems 
that are used in practice is not known. Therefore, the universality of the 
distribution functions of the computation time and other quantities related
to computational complexity is of great importance. It would imply that 
distributions found for the idealized systems may be relevant 
also for the realistic sets of LP problems. In this paper we demonstrate 
numerically that for several probability distributions of LP problems, 
the functions of the quantities that characterize the complexity
are indeed universal, providing support for the conjecture that universality 
holds in general. 

It was shown in \cite{toda} that the flow equations corresponding to 
(\ref{field}) are, in fact, part of a system of Hamiltonian equations of 
motion of a completely integrable system of a Toda type. Therefore, like the 
Toda system, it is integrable with the formal solution \cite{faybusovich}
\begin{equation}
\label{solution}
x_i(t) = x_i(0) \exp \left( -\Delta_i t +
\sum_{j=1}^{m} \alpha_{ji} \log \frac{x_{j+n-m}(t)}{x_{j+n-m}(0)} \right)
\end{equation}
($i = 1,\ldots ,n-m$), that describes the time evolution of the $n-m$
independent variables $x_{\cal N}(t)$, in terms of the variables
$x_{\cal B}(t)$. In (\ref{solution})
$x_i(0)$ and $x_{j+n-m}(0)$ are components of the initial
condition, $x_{j+n-m}(t)$ are the $x_{\cb}$ components of the solution,
$\alpha_{ji}= - (B^{-1} N)_{ji}$
is an $m \times (n-m)$ matrix, while
\begin{equation}\label{deltas}
\Delta_i = -c_i  - \sum_{j=1}^{m} c_j \alpha_{ji}\,.
\end{equation}
For the decomposition
$x=(x_{{\cal N}},x_{{\cal B}})$
used for the optimal vertex, $\Delta_i \geq 0~~i=1,\ldots,n-m\,,$
and $x_{\cal N}(t)$ converges to 0, while
$x_{\cal B}(t)$ converges to $x^*=B^{-1}b$.
Note that the analytical solution is only a {\em formal} one, and does not
provide an answer
to the LP instance, since the $\Delta_i$ depend on the partition of $A$, and
only relative to a partition corresponding to a
maximum vertex are all the $\Delta_i$ positive.

The second term in (\ref{solution}),
when it is positive, is a kind of ``barrier'':
$\Delta_{i}t$
must be larger than the barrier before $x_i$ can decrease to zero.
In the following we ignore the contribution of the initial condition
and denote the value of this term in the infinite time limit by
\begin{equation}
\label{barrier}
\beta_i = \sum_{j=1}^m \alpha_{ji} \log x_{j+n-m}^*.
\end{equation}
In order for $x(t)$ to be close to the maximum vertex we must have
$x_i(t) < \epsilon$ for $i=1,\ldots,n-m$ for some small positive
$\epsilon$, namely
$\exp (- \Delta_i t + \beta_i) < \epsilon ~,~~ \mbox{for}~ i =
1,\ldots,n-m.$
Therefore we consider
\begin{equation}\label{T}
T = \max_{i} \left( \frac{\beta_i}{\Delta_i} +
    \frac{|\log \epsilon |}{\Delta_i} \right)~,
\end{equation}
as the computation time.
We denote
\begin{equation}\label{Deltamin}
\Delta_{\min} = \min_i \Delta_i,~~~\beta_{\max} = \max_i \beta_i \;.
\end{equation}
The $\Delta_i$ can be arbitrarily small when the inputs are real numbers. 
Such ``bad'' instances (associated with the possibility of vanishing 
$x_{j+n-m}^*$'s) are rare in the Gaussian probabilistic model of 
\cite{bffs, pla}, where it was shown that typically 
$\Delta_{\min}\sim 1/\sqrt{m}$. In this Letter we will 
show that this scaling behavior of $\Delta_{\min}$ is universal, being valid 
in a broad  class of probability distributions of LP problems. 


Consider an ensemble of LP problems in which the components of $(A,b,c)$ are 
independent identically distributed (i.i.d.) random variables taken from
various {\em even} distributions, with 0 mean and bounded variance.
For a probabilistic model of LP instances,
$\Delta_{\min},\, \beta_{\max}$ and $T$ are random variables.

In \cite{bffs, pla}, the components of $A$, $b$, and $c$ 
were taken from the Gaussian distribution (see, e.g., Eqs.(12-18) 
in \cite{bffs}) with zero mean and variance $\sigma^2$. It was found 
analytically, in the large 
$(n,m)$ limit, that the probability ${\cal P}(\Delta_{min} < \Delta|
\Delta_{min} > 0) \equiv {\cal F}^{(n,m)}(\Delta)$
is of the scaling form
\begin{equation} \label{scaling.delta}
{\cal F}^{(n,m)}(\Delta)=1-e^{x_\Delta^2}\,{\rm erfc}(x_\Delta)\ \equiv {\cal
F}(x_\Delta)
\end{equation}
with the scaling variable 
\begin{equation}
x_\Delta = a_\Delta\left(n/m\right) x_\Delta'\,,
\label{xd}
\end{equation}
where
\begin{equation}
 x_\Delta'  = \sqrt{m} \Delta\quad {\rm and}\quad
a_\Delta\left(n/m\right)  = \frac{1}{\sqrt{\pi}} (\frac{n}{m}-1) 
\frac{1}{\sigma}\,.
\label{ad}
\end{equation}
The scaling function ${\cal F}$ contains {\em all} asymptotic information on 
$\Delta$. The distribution ${\cal F}(x_\Delta)$ is very wide and
does not have a variance. Also the average of $1/x_\Delta$ diverges.

The amazing point is that in the limit of large $m$ and $n$, the probability 
distributions of $\Delta_{\rm min}$ depend on the variables $m$, $n$ and 
$\Delta$ only via the scaling variable $x_\Delta$. 
If the limit of infinite $m$ and $n$ is taken, so that $n/m$ is fixed, 
$a_\Delta $ is constant. It was verified numerically that for the 
Gaussian ensemble (\ref{scaling.delta}) is a good approximation already for 
$m=20$, and $n=40$.

The existence of scaling functions like (\ref{scaling.delta}) for the
barrier $\beta_{max}$ (that is, the maximum of the $\beta_i$) defined by 
(\ref{barrier}), and for  $T$ defined by (\ref{T}) (assuming that $\epsilon$ 
is not too small so that the first term in (\ref{T}) dominates), was verified 
numerically for the Gaussian distribution \cite{bffs, pla}. 

The scaling behavior (\ref{scaling.delta}), and similar behavior found for 
$1/\beta_{max}$ and $1/T$, rendering their distribution functions 
$P_\beta(1/\beta)$ and $P_T(1/T)$ scaling functions of the scaling variables
$x_\beta$ and $x_T$, all associated with the Gaussian ensemble, prompted us, 
for reasons discussed above, to explore their universality and check their 
validity for other probability ensembles of LP problems. Thus, we carried 
numerical calculations of the distribution functions of $\Delta$, $\beta$, 
and $1/T$ for various probability distributions of $A$, $b$, and $c$. 

To be specific, we studied: (1) the uniform distribution, in which each entry 
of $(A,b,c)$ was uniformly distributed between $\pm\frac{1}{2} $; (2) the 
discrete, bimodal distribution, in which each entry was either +1 or -1 with 
probability $\frac{1}{2} $ each; and finally, (3) the diluted bimodal 
distribution, in which each entry was either +1 or -1 with probability 
$\frac{p}{2}$ each, or 0 with probability $1-p$. Here we chose $p=0.2, 0.5$
in numerical calculations.

For continuous distributions the probability of encountering degenerate 
solutions (where some of the $x_{j+n-m}^*$'s in (\ref{barrier}) may vanish)
is of measure zero, while for the discrete ensembles some regularization
was introduced, in order to avoid degenerate situations.

We generated full LP instances $(A,b,c)$ with the probability 
distribution in question. For each instance the LP problem was solved using 
the linear programming solver of MatLab. Only instances with a bounded optimal 
solution were kept, and $\Delta_{\min}$ was computed relative to the optimal 
partition and optimality was verified by checking that $\Delta_{\min} >0$.
Using the sampled instances we obtain an estimate of 
\begin{equation}\label{universalFDelta}
{\cal F}^{(n,m)}(\Delta) \equiv P_\Delta (\Delta) = 
{\cal P}(\Delta_{min} < \Delta |\Delta_{min} > 0)
\end{equation}
and of the corresponding cumulative distribution functions of the barrier 
$\beta_{\max}$ and the computation time $T$.

The solution of the  LP problem is used here in order to identify the optimal 
partition of $A$ into $B$ and $N$. This enables one to compute 
$\Delta_{min}$, $\beta_{max}$, and $T$ from (\ref{T}) and (\ref{Deltamin}), 
and the distribution $P_\Delta(\Delta)$ as well as $P_\beta(1/\beta)$ 
and $P_T(1/t)$.

\begin{figure}[htb]
\epsfig{file=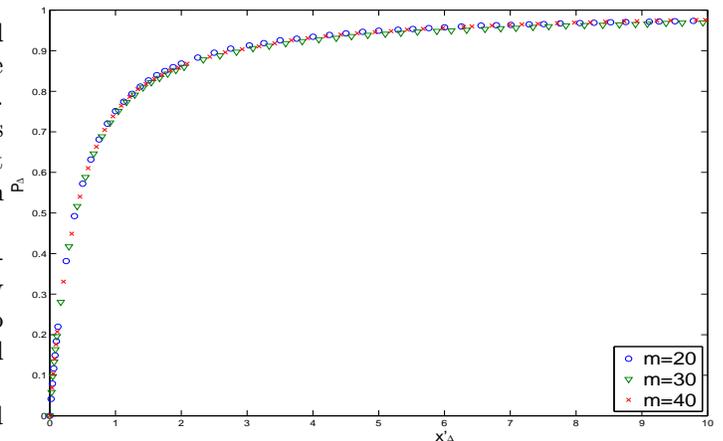,height=8cm,width=7cm,angle=0}
\vspace{-2.0cm}
\caption{${\cal P}_\Delta$ is plotted as a function of 
$x'_\Delta$ for the uniform distribution for LP problems with $n=2m$. The 
number of instances used in the simulation is 121939 
for the $m=20$ case, 91977 for the $m=30$ case and 112206 for the $m=40$ 
case. The number of converging instances for each case is 20000.
\label{delta-uniform}}
\end{figure}

For example, the scaling behavior of (\ref{universalFDelta}) in the case of 
the uniform ensemble can be seen in Fig. \ref{delta-uniform}.  
Similar scaling behavior is found also for $P_\beta$ and $P_T$ for the uniform 
distribution. Scaling behavior of this nature was confirmed also for the 
bimodal distribution, and preliminary results suggest that it holds also for 
the diluted bimodal distribution.

A natural question which arises is whether the distribution functions 
$P_\Delta$, $P_\beta$ and $P_T$ are universal \cite{bffs, pla}. 
In other words, do all probability ensembles of LP problems, or at least a 
large family thereof, yield the same functions $P_\Delta$, $P_\beta$ and 
$P_T$ of the scaling variables $x_\Delta, x_\beta$ and $x_T$?
We found that the answer to this question is on the affirmative. 

\begin{figure}[htb]
\epsfig{file=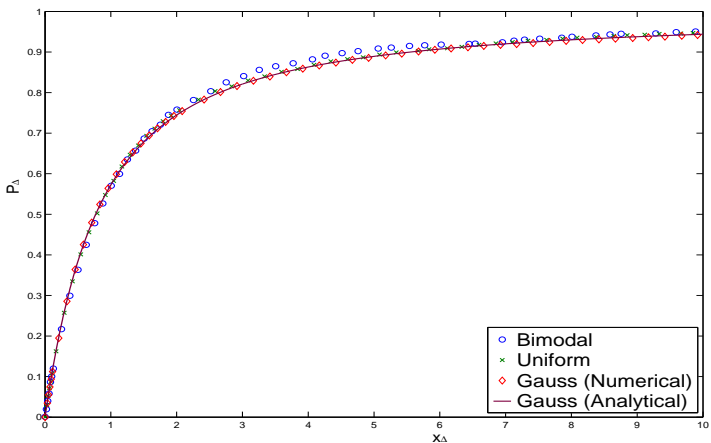,height=8cm,width=7cm,angle=0}
\vspace{-2.0cm}
\caption{ $P_\Delta$ as a function of $x_\Delta$. Here $n=2m$ and $m=40$. 
The scale factors $a_\Delta$ are: $1.044/\sqrt{\pi}$ for the bimodal 
distribution, $3.603/\sqrt{\pi}$ for the uniform distribution, 
and $1/\sqrt{\pi}$ for the gaussian distribution. For the Gaussian 
and bimodal distributions the variance is $\sigma^2=1\,,$ while for the 
uniform distribution $\sigma^2=1/12\,.$\label{14}}
\end{figure}

Specifically for $\Delta$, plots like Fig. \ref{delta-uniform} were 
produced for the variables $x'_\Delta$ for the various distributions that were
studied. Indeed, these were found to be scaling functions. Then, the scale
factors $a_\Delta$, corresponding to (\ref{ad}) were calculated for the 
various distributions so that $P_\Delta$ as a function of $x_\Delta$ is the 
same function for all distributions. (This is done, as usual, by least
squares fit.) Indeed, a universal function for $P_\Delta$ is found, as 
is clear from  Fig. \ref{14}, and it reduces to (\ref{scaling.delta}) 
that was found for the Gaussian ensemble in \cite{bffs, pla}.
It turns out that all the scaling factors that were found in this way for 
the various distributions are in agreement with (\ref{ad}) (in the way it 
depends on $\sigma$). Similar scaling was found also for the distribution
functions $P_\beta$ and $P_T$ of $1/\beta_{max}$ and $1/T$, respectively.
        
In summary, we find that the asymptotic distribution functions 
$P_\Delta, P_\beta$ and $P_T$ are the same for all distributions of the 
parameters studied here, if expressed in terms of the scaling 
variables $x_\Delta, x_\beta$ and $x_T$. Furthermore, we find that then the 
following combinations of scaling factors, $\sigma a_\Delta, 
a_\beta$ and $\sigma a_T$ are independent of the 
probability distribution. Therefore, in particular, $a_\Delta$ should 
satisfy (\ref{ad}) that was found for the Gaussian distribution, 
while $ a_\beta$ and $a_T$ are yet to be found analytically.

Based the results presented in this paper, as well as the results of 
\cite{bffs, pla}, we conjecture that the scaling behavior of the various 
distribution functions (and the corresponding scaling factor) is {\em 
universal}, i.e., that it is robust and should be valid in a large class of 
ensembles of LP problems, in which the $(A,b,c)$ data are taken from 
even distributions, with zero mean and finite variance.

\begin{acknowledgments}
It is a pleasure to thank our colleagues Asa Ben-Hur and Hava Siegelmann 
for advice and discussions. This research was supported in part by the Shlomo 
Kaplansky Academic Chair, by the Technion-Haifa University Collaboration Fund,
by the US-Israel Binational Science Foundation (BSF), by the Israeli Science 
Foundation, and by the Minerva Center of Nonlinear Physics of Complex Systems.
\end{acknowledgments}

\typeout{References}

\end{document}